\documentclass[12pt]{iopart}
\usepackage{bm}
\usepackage{graphicx}
\usepackage{cite}
\usepackage{color}
\usepackage{siunitx}
\usepackage{color} 
\usepackage{cancel}
\usepackage{threeparttable}
\expandafter\let\csname equation*\endcsname\relax
\expandafter\let\csname endequation*\endcsname\relax
\usepackage{amsmath}

\begin{document}
\title{Topological end state and enhanced thermoelectric performance of a supramolecular device}

\author{Wenlai Mu$^{1}$, Nisar Muhammad$^{2}$, Huaihong Guo$^{3,*}$,Zsolt Gulacsi$^{4}$, Teng Yang$^{1,*}$, Zhidong Zhang$^{1}$}
\address{$^1$Shenyang National Laboratory for Materials Science, Institute of Metal Research, Chinese Academy of Sciences, School of Materials Science and Engineering, University of Science and Technology of China, Shenyang 110016, China}
\address{$^{2}$Hefei National Laboratory for Physical Sciences at Microscale and Department of Physics, University of Science and Technology of China, Hefei 230026, China}
\address{$^{3}$College of Sciences, Liaoning Petrochemical University, Fushun 113001, China}
\address{$^{4}$Department of Theoretical Physics, University of Debrecen, Debrecen, Hungary}
\ead{hhguo@alum.imr.ac.cn; yanghaiteng@msn.com}

\vspace{10pt}
\begin{indented}
\item[]Received 1 January 2025
\end{indented}
\vspace{2pc}
\begin{abstract}
Supramolecular device (SMD) with topological end states and a noncovalent junction is rarely investigated but deemed promising for thermoelectric (TE) applications. We designed a new kind of SMD based on the Su-Schrieffer-Heeger (SSH) chains, and calculated TE properties of it using the non-equilibrium Green's function (NEGF) method. By scaling TE performance under different optimization conditions, we found the best scenario. Our result shows that the existing topological end states indeed give rise to a large value of power factor, rendering a dimensionless figure-of-merit ZT above 2 in a broad range of chemical potential (doping). Moreover, by imposing the system to various perturbations including end state shift, structural change and disorder, we found that the SMD system possesses a prominent switch effect, further optimizing its performance for TE applications. 
\end{abstract}
\noindent{\it Keywords}: Su-Schrieffer-Heeger model, topological end state, non-equilibrium Green’s function method, polyacetylene, thermoelectric supramolecular device
\pacs{72.20.Pa} \maketitle

\bigskip

\section{Introduction}
Thermoelectric (TE) devices have recently attracted much attention due to their potentially high efficiency of energy conversion between heat and electricity~\cite{Tedevice,Tedevice2}. The performance of thermoelectric (TE) devices is usually evaluated using the dimensionless figure of merit $ZT=\mathcal{G} S^2 T/(K_e+K_r)$, in which $\mathcal{G}$ denotes the electrical conductance, $K_e$ the electronic thermal conductance, $K_r$ the thermal conductance beyond $K_e$, and $S$ the Seebeck coefficient which is defined by $S=-d\phi^{e}/dT$ with $\phi^{e}$ as electric potential. To achieve a high value of $ZT$ and flexibility, research on TE devices has extended to the nanoscale, leading to the emergence of numerous new categories~\cite{nano1,nano2,nano3}, such as the molecular device (MD)~\cite{md1,md2,md3}, a popular type of nanodevices for TE applications~\cite{mdte1,mdte2,mdte3}. Previous investigations of molecular devices were largely focused on single molecular devices; nevertheless, covalent bonds in the structure of devices are detrimental to their efficiency for TE applications due to their good thermal transport property~\cite{mdtc}. Recently, increasing attention has been drawn to molecular junctions without covalent connections~\cite{smdte1,smdte2,smdte3}. The molecular device possessing these characteristics is known as the supramolecular device (SMD)~\cite{smd}. Because of the noncovalent coupling between the device constituents, SMD provides a platform with high flexibility and promising TE application because of the strong phonon scattering in the noncovalent device junctions. 

In this circumstance, the following questions may arise. What if the topological end state is introduced into the SMD? Can we further optimize the TE properties of SMD? The effect of topological boundary states on TE performance has already been studied since the discovery of topological insulators (TI)~\cite{tpte1,tpte2,tpte3,tpte4,tpte5,hfc}. TIs tend to have a significant Seebeck effect with high conductivity to become an effective TE material. Previous investigations were mainly on three- and two-dimensional TIs, 2D TIs with edge states seem to possess better TE performance than their 3D counterpart~\cite{2dtpte1,2dtpte2}, indicating that reduced dimensionality is beneficial for TE performance. Therefore, it is curious to wonder if one-dimensional topological material with topological boundary state (\textit{i.e.} topological end state) in TE devices can work as well.  

\begin{figure*}
\begin{center}\scalebox{0.30}{\includegraphics{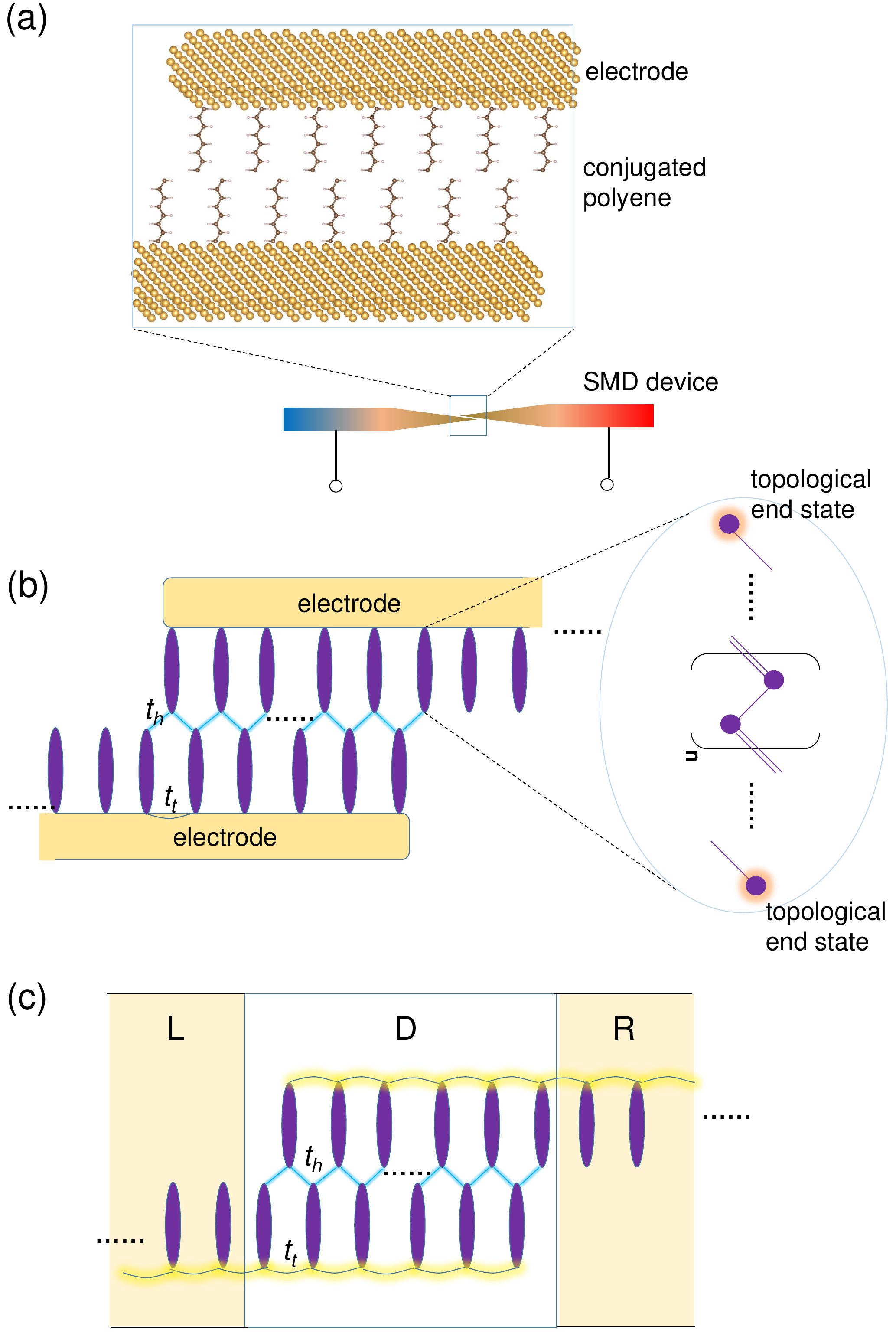}}
\caption{Supramolecular device (SMD). (a) The possible real and (b) schematic SMD made of the SSH chains. The device system consists of two electrodes with one end of SSH chains adsorbed on the electrode surface and the other end crisscross connected with each other via non-covalent connections. SSH chain is zoomed in (b) to show the topological end states.  (c) Schematic of SMD module for non-equilibrium Green's function (NEGF) calculations. 
\label{fig1}}
\end{center}
\end{figure*}

To study the effect of topological end state on TE properties of SMD, the Su-Schrieffer-Heeger (SSH) model, which was first introduced to describe the polyacetylene single molecular system~\cite{ssh1,ssh2}, was used. The SSH model could possess end states under open boundary condition (OBC) if it is in a topological non-trival phase. The Hamiltonian of this model is given by 
\begin{equation}\label{SSH}
\begin{split}
&H_{SSH}=\sum_{i,\alpha}\varepsilon_{i,\alpha}c_{i,\alpha}^{\dagger}c_{i,\alpha} \\
&+\sum_{<ij>}(t_{1}c_{i,\alpha}^{\dagger}c_{i,\beta}+t_{2}c_{i,\alpha}^{\dagger}c_{j,\beta}+h.c.) ,
\end{split}
\end{equation}
where $c_{i,\alpha}$ represents the annihilation operator of an electron on the $\alpha$ sublattice of the $i^{th}$ unit cell. Under periodic boundary condition (PBC), the homotopy invariant of such a system can be described by the winding number of its bottom manifold, which is defined as
\begin{equation}\label{W}
\begin{split}
W&=\frac{\oint_{BZ} \mathbf{A} \cdot d\mathbf{k}}{\pi}=\frac{\oint_{BZ} <i\partial_k> \cdot d\mathbf{k}}{\pi} \\
&=\frac{1-sgn(2\delta)}{2} , 
\end{split}
\end{equation}
where $\delta=\frac{t_{2}-t_{1}}{2}$ is the energy difference between the inner-cell hopping ($t_{1}$) and inter-cell hopping ($t_{2}$). 
When $t_{2}$ is stronger than $t_{1}$, \textit{i.e.} $t_{2} < t_{1} < 0$, the system is topological nontrivial and possesses topological end states under OBC.~\cite{tpssh} To highlight the significance of topological end states in TE optimization, Lima \emph{et al.} designed a TE device based on topological SSH chain and quantum dot~\cite{sqs}. A maximum ZT value of up to 30 is achieved, but external thermal conductance is not considered. 

A direct approach to realize SSH chains is using polyacetylene, which is light-weighted and flexible in nature and widely explored in wearable and molecular devices~\cite{expssh}. More importantly,
the incredible electrical conducting ability of polyacetylene upon doping~\cite{condpoly} makes it promising for thermoelectric application.

In this work, we first designed a new SMD structure with noncovalent connections based on the SSH chain, and then investigated its TE performance using non-equilibrium Green’s function (NEGF) method. 
The topological phase transition and the end states of the SSH chain unit in the SMD are found to have a significant impact on the electrical transport property. The maximum ZT value of the SMD system can reach 5 in a wide range of chemical potential (doping), after external thermal conductance is practically taken into account.~\cite{gaffar2021effects} Moreover, the SMD system is found to possess a prominent switch effect, further optimizing its performances for thermoelectric device applications. 

\section{Model and Method}
Based on SSH chain, we constructed a nanowire superstructure, as shown in Fig.~\ref{fig1}. 
The Hamiltonian of the system can be written as $H=\sum_{n}H_{SSH,n}+\sum_{\mu,\nu}H_{\mu \nu}$, with $n$ denoting the $n^{th}$ SSH chain unit. The second term $H_{\mu \nu}$ represents the coupling between the $\mu$th and $\nu$th SSH chains, and is expressed as
\begin{equation}\label{T}
\begin{split}
H_{\mu \nu}=&t_{h}\sum_{\mu,\nu}c_{h,\mu}^{\dagger}c_{h,\nu} +t_{t}\sum_{\mu,\nu}c_{t,\mu}^{\dagger}c_{t,\nu} \\
&+h.c. ,
\end{split}
\end{equation}
$t_{h}$ and $t_{t}$ are the hopping energies between heads and between tails of SSH chain, respectively, as given in Figs.~\ref{fig1} (b) and (c). 
In this work, we focus on the region of $t_{0}\leq t_{h(t)}<0$. $t_{0}=\frac{t_{2}+t_{1}}{2}$ is the average hopping energy in an SSH chain. To numerically illustrate the TE properties, we take the data of polyacetylene for the parameters, the value of $t_{0}=-3.002$ eV and $\delta = \frac{t_{2} - t_{1}}{2} = 0.028\lambda |t_{0}|$ is taken from the previous work\cite{sshtb}, in which $\lambda=\pm 1$ represents the topological phase of unit SSH chains in the system, \textit{i.e.} $\lambda <= 0$ ($\lambda > 0$) signifies topological non-trivial (topological trivial). 
Moreover, according to the previous work\cite{sshtb}, the parameter $t_{h}$ depends on the distance $d$ between the heads of neighboring polyacetylene SSH chains, namely, $t_{h}=a_1-b_1(d-d_0)/d_0$, where $a_1=0.82t_0$, $b_1=0.98t_0$, and $d_0=1.48$ Å. $t_{h}$ changes from $t_0$ to 0 as one tunes $d$ from 1.21 Å to 2.72 Å.

The transport and TE properties were calculated via linear-response theory based on the non-equilibrium Green's function (NEGF) formalism~\cite{qd}. In the linear-response regime, conductance of a device is given by 
\begin{equation}\label{G}
\mathcal{G}=\lim_{V\to 0}\frac{dJ_{e}}{dV}=\frac{e^{2}}{h}L_{0} ,
\end{equation}
where $h$ is Planck's constant, $L_{0}$ is the zero order of the transport integral, which can be expressed as
\begin{equation}\label{Kn}
L_n=\int(E-\mu)^n(-\frac{\partial f}{\partial E})\mathcal{T}(E)dE ,
\end{equation}
where $\mu$ is the chemical potential, $f=1/(e^{(E-\mu)/k_BT}+1)$ is the Fermi distribution function, $k_B$ is Boltzmann's constant, and $\mathcal{T}(E)$ is the transmission spectral function with energy as variable. By applying linear-response theory on TE process in devices, we can write the current vector up to the linear terms of potential difference,
\begin{equation}\label{J}
\begin{aligned}
   J_{\gamma}=\mathcal{L}_{\gamma \zeta}(\phi^{\zeta}_R-\phi^{\zeta}_L) ,
\end{aligned}
\end{equation}
where $J$ is the current, $\phi$ is the potential, indexes $\gamma, \zeta=e, q$ denote the electrical and thermal part, respectively. 
For electrical transport, $\phi^{e}_R-\phi^{e}_L=V$ is the voltage between two leads. For thermal transport, $\phi^{q}_R-\phi^{q}_L=T_R-T_L=T_{\Delta}$ is the temperature difference between two leads, and $\frac{\phi^{q}_R+\phi^{q}_L}{2}=\frac{T_R+T_L}{2}=T$ is the average temperature of the system. The coefficients are in relation with the transport integral, such as $\mathcal{L}_{ee}=e^{2}L_{0}/h$, $T\mathcal{L}_{eq}=\mathcal{L}_{qe}=-eL_{1}/h$, $\mathcal{L}_{qq}=L_{2}/hT$. 
Accordingly, the electronic thermal conductance
\begin{equation}\label{kap}
K_{e}=\lim_{T_{\Delta}\to 0}\frac{dJ_{q}}{dT_{\Delta}}=\frac{1}{T}(L_{2}-\frac{(L_{1})^{2}}{L_{0}}) ,
\end{equation}
and Seebeck coefficient
\begin{equation}\label{S}
S=-\frac{d\phi^{e}}{dT}|_{J_{\gamma}=0}=-\frac{L_{1}}{eTL_{0}} .
\end{equation} 
For a realistic consideration, an extra thermal conductance $K_r$ in unit of $K_0$ ($K_0=1\times 10^{-10}$ W/K, a reference value of thermal conductance for polyacetylene materials at 300 K)~\cite{patc} was introduced. 

$\mathcal{T}$ in equation~\eqref{Kn} can be written as
\begin{equation}\label{tau}
\mathcal{T}=\mathrm{Tr}[\Gamma_{L}G\Gamma_{R}G^{\dagger}] ,
\end{equation}
where $\Gamma_{L(R)}=i(\Sigma_{L(R)}-\Sigma^{\dagger}_{L(R)})$. $\Gamma_{L(R)}$ and $\Sigma_{L(R)}$ are the broadening function and the self energy of contacts, respectively.  
$G$ is the retarded Green's function of the device, which under the energy representation can be expressed as
\begin{equation}\label{Green}
 G(E)=[EI-H-\Sigma_{L}-\Sigma_{R}]^{-1}  .
\end{equation}
According to the Sancho-Rubio iterative scheme~\cite{Sancho85}, self-energy matrices of contact can be presented as
\begin{equation}\label{self en}
\Sigma_{L(R)}=H_{DL(R)}g_{L(R)}H_{L(R)D} ,
\end{equation}
where $g$ is the surface Green's function, D, L and R denote the Hamiltonian in the D, L and R.  The total Hamiltonian can be written as
\begin{equation}\label{HLDR}
{H}={ \left[
 \begin{matrix}
   H_{LL} & H_{LD} & 0 \\
   H_{DL} & H_{DD} & H_{DR} \\
   0 & H_{RD} & H_{RR}
  \end{matrix}
  \right]} ,
\end{equation}
in which $H_{LL}=H_{RR}=\sum_{n}H_{SSH,n}+t_{t}\sum_{\mu,\nu}(c_{t,\mu}^{\dagger}c_{t,\nu}+h.c.)$ are the Hamiltonian for leads, $H_{DD}=\sum_{n}H_{SSH,n}+\sum_{\mu,\nu}H_{\mu \nu}$ for device, where $H_{SSH,n}$ are from Eq.~\eqref{SSH} and $H_{\mu \nu}$ are from Eq.~\eqref{T}, $H_{DL}$ = $t_{t}(c_{t,D}^{\dagger}c_{t,L}+h.c.)$ is the Hamiltonian for hopping between D and L. 
The boundary conditions for $H_{LL}$ and $H_{RR}$ are chosen as semi-boundless. 

By incorporating the Hamiltonians within the NEGF formalism, one can achieve the Green function via iterative algorithm~\cite{Sancho85}, and the TE properties of the system.

\begin{figure*}
\begin{center}\scalebox{0.45}{\includegraphics{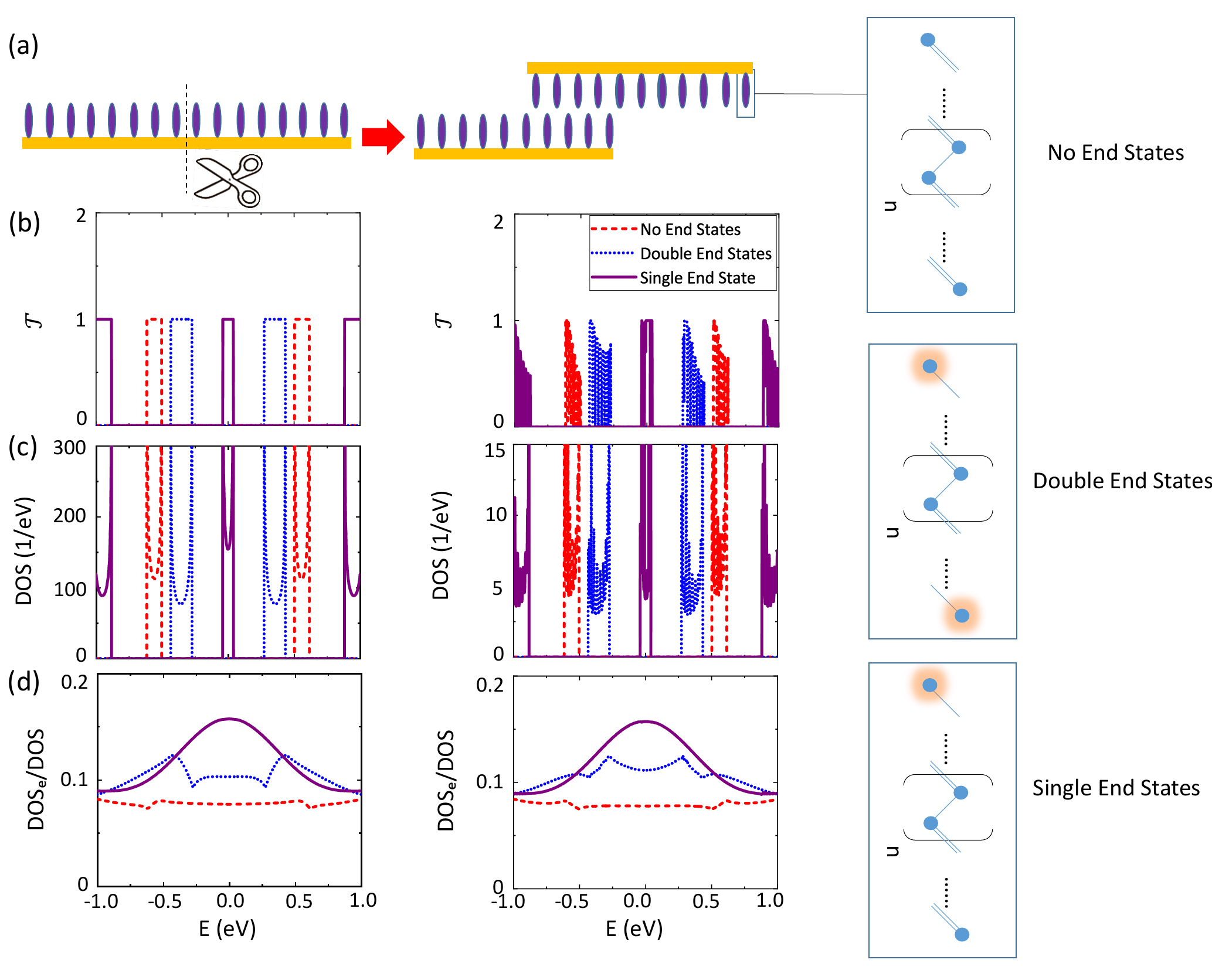}}
\caption{Structure dependent electronic properties. (a) The super SSH chain (left) and SMD (right) structures. Purple ellipsoid denotes SSH chain, and the yellow band denotes the electrode connection. 
(b) Transmission spectral function $\mathcal{T}$. (c) Total density of states (DOS). (d) Ratio between DOS on the end of SSH chain and total DOS. In (b-d), the $ne$, $de$ and $se$ cases are represented by red dot line, blue dash line and purple solid line, respectively. 
\label{fig2}}
\end{center}
\end{figure*}

\section{Numerical results and discussion}

\subsection{Transmission spectrum and density of states}
Transmission spectrum and density of states of the system are illustrated in this section to provide an original image about electric properties of the SMD designed before studying its TE properties.

\begin{figure*}
\begin{center}\scalebox{0.38}{\includegraphics{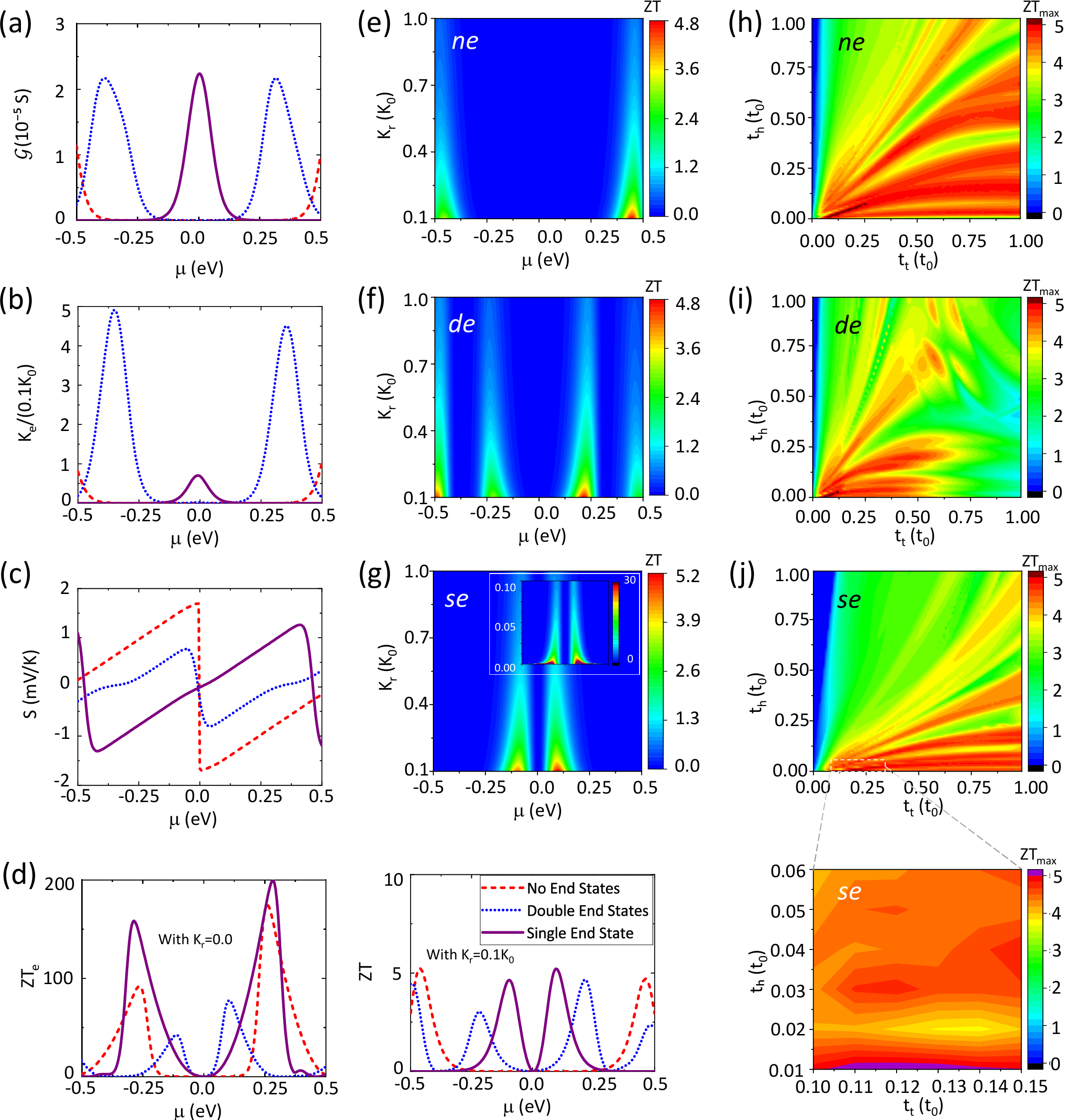}}
\caption{Thermoelectric properties of the SMD structure in the three end states. Chemical potential $\mu$ dependence of (a) electrical conductance $\mathcal{G}$, (b) electronic thermal conductance $K_e$, (c) Seebeck coefficient $S$, and (d) figure of merit $ZT$ value. The $ne$, $de$ and $se$ cases are respectively represented by red dot line, blue dash line and purple solid line. In (a,c), only electronic thermal conductance $K_e$ is considered. $ZT$ with external thermal conductance $K_r$ = 0 and $0.1K_0$ are given on the left and right hand side of (d), respectively. $ZT$ value as a function of external thermal conductance $K_r$ and chemical potential $\mu$ are given in the (e) $ne$, (f) $de$ and (g) $se$ cases. The maximal $ZT$ value as a function of inter SSH hopping energy $t_t$ and $t_h$ are given in the (h) $ne$, (i) $de$ and (j) $se$ cases. The optimal zone [$t_h=0.01t_0$ and $t_t=0.12t_0$] for the $se$ case is marked by white dashed square on top panel of (j) and is zoomed in on bottom panel of (j). Temperature is set at 300 K. 
\label{fig3}}
\end{center}
\end{figure*}

Practically, we designed a non-covalent connection by dividing one infinite long chain (super SSH chain) into two semi-infinite chains (SMD), as shown in Fig.~\ref{fig2} (a), to inhibit phonon transport. Various end structures can be constructed based on SSH chain, as shown on the right-hand side of Fig.~\ref{fig2} (a). The ends of the SSH chains were connected so that the function of end states could be manifested. By tuning the number of atoms in the SSH chain, the chain can possess different topological phases, either no end state (marked as $ne$ case), or double end state (marked as $de$ case), or single end state (marked as $se$ case). 
To note that when the winding number $W$ of a SSH chain is zero, it is in the topological trivial phase or the $ne$ state. While it is in the topological non-trivial phase when $W=1$, either in the $de$ state when it has a center symmetric boundary, or in the $se$ state when the center symmetry of the boundary is broken, \textit{i.e.} the number of atoms is odd. The onsite energy is taken to be zero, and the number of atoms is set to 20 (19 for the $se$ case), and the inter SSH chain hopping energy $t_h$ and $t_t$ here are set as $t_h=0.01t_0$ and $t_t=0.12t_0$ (the optimal condition for the $se$ case of the system, as shown in Fig.~\ref{fig3} (d-f), and more details given in Section.~\ref{CO}).

We here demonstrate the impact of various end states on electronic properties of the SSH structure. In Figs.~\ref{fig2} (b) and (c), the spectrum of transmission function and density of states (DOS) are presented. For an infinitely long wire, as can be observed on the left panel of Fig.~\ref{fig2}, the system with no end states (the $ne$ case) possesses a relatively large energy gap. After introducing double end states (the $de$ case), the gap will shrink as a result of topological end states. If a single end state (the $se$ case) is concerned, the end states of the system will appear around the Fermi level. In this case, the system will exhibit a single-band structure feature near Fermi level. The end states at the Fermi level are confirmed by the local DOS, as presented in Fig.~\ref{fig2} (d). Here, the super SSH chain in the $se$ case is expected to possess TE performance due to its single-band characteristic, in contrast to the $ne$ and $de$ cases, both of which exhibit a level splitting band gap.  


We further demonstrate the impact of non-covalent connection on the electronic properties of the SMD structure. The non-covalent connection is via van der Waals interactions between the ends of two SSH chains. 
Compared to its periodic counterpart on the left panel of Fig.~\ref{fig2}, the SSH chain SMD on the right has its transmission much attenuated and an antiresonance-like fluctuation in the spectrum~\cite{fano} in both $ne$ and $de$ cases. However, electronic properties in the $se$ case is almost unaffected. Retaining almost perfect transmission around the Fermi level, the $se$ case in the SMD structure still has end states gathered around the Fermi level and the transmission step much less perturbed by the non-covalent connection, showing topological robustness. 

\subsection{Thermoelectric properties and conditions of optimization} \label{CO}

In this section, we continue to investigate the thermoelectric (TE) properties of the three cases of the end states. We will find that the $se$ case stands out with an optimal TE performance among the three cases. 

The chemical potential $\mu$ dependent electrical conductance $\mathcal{G}$, electronic thermal conductance $K_e$, Seebeck coefficient $S$ and ZT value are shown in Figs.~\ref{fig3} (a-d). Here an optimal condition of $t_h=0.01t_0$ and $t_t=0.12t_0$, as will be demonstrated in the upcoming Fig.~\ref{fig3} (j), is chosen. The chemical potential is fixed in the range of [-0.5 eV, 0.5 eV], since the peak performance appears in this region. 
From Figs.~\ref{fig3} (a,b), the number of end states significantly affects the electrical conductance and electronic thermal conductance in the vicinity of $\mu$ = 0. Both $ne$ and $de$ cases exhibit a gap with two symmetrically distributed peaks around $\mu$ = 0. The $de$ has a smaller gap than the $ne$. However, for the $se$ case, a single conductance peak emerges symmetrically around $\mu$ = 0. As the end states gathered at the junction form a quantum dot-like structure, it is no wonder that the conductance characteristic of the $se$ case resembles a single quantum dot device~\cite{qd}. It is also interesting to see that both $\mathcal{G}$ and $K_e$ have peak values at the same chemical potentials for all three cases, which may be understood by the Wiedemann-Franz Law. However, the ratio of $\frac{K_e}{\mathcal{G}T}$, which is supposed to hold a Lorenz number constant as the Wiedemann-Franz Law requires, seems to change from case to case here. Among the three cases, the $se$ has the largest value of $\frac{\mathcal{G}}{K_e}$, which has direct relevance to $ZT$ and suggests its advantage over the other two cases in the thermoelectric properties.


Seebeck coefficients $S$, as another important parameter for evaluating the TE performance, are shown in Fig.~\ref{fig3} (c). The chemical potential dependence of $S$ follows the Mott formula, namely, $S(\mu) \sim -\frac{\partial \mathcal{G}}{\partial \mu}$. This will lead to a chemical potential mismatch between the peak values of $S$ (Fig.~\ref{fig3} (c)) and $\mathcal{G}$ (Fig.~\ref{fig3} (a)). In Fig.~\ref{fig3} (c), the $S$ behaviors of the three cases are rather distinct from each other. For both the $ne$ and $se$ cases, Seebeck coefficient diverges almost linearly in the center of its large gap, reaching as high as around 1 mV/K. While, for the $de$ case, a similar trend can be observed, but $S$ does not diverge and acts more like a bulk semiconductor. 





The divergence of $S$ is advantageous for the dimensionless figure of merit $ZT$. The $\mu$-dependent $ZT_e$ on the left side of Fig.~\ref{fig3} (d) shows that both $ne$ and $se$ cases share a similar large magnitude of peak $ZT_e$ values, up to 200, around $\mu=0.25$ eV. This is much higher than previously reported, which is simply based on SSH-chain-related molecular device~\cite{sqs}. The $de$ case has the lowest $ZT_e$ value. 

For realistic applications, external thermal conductance $K_r$ (i.e. due to phonon) in addition to electronic counterpart is too inevitable to neglect. Since $ZT$ = $ZT_e$ $\frac{1}{1 + K_r/K_e}$, one expects that the peak positions should be shifted from $ZT_e$ to $ZT$ due to the external thermal conductance $K_r$ and $K_e$ in the second term. Indeed, as illustrated in Figs.~\ref{fig3} (d), from left ($K_r$ = 0) to right ($K_r = 0.1K_0$), one can see that $ZT$ peak are all shifted. Depending on the relative peak positions of both $ZT_e$ and $K_e$, the $se$ case has its $ZT$ peaks shifted to smaller value of $\mu$, while both the $de$ and $ne$ cases have their $ZT$ peaks shifted to larger $\mu$, making the $se$ case more preferable in terms of the small doping level.  



To optimize the combination of tuning parameters for decent TE performance, we here investigate the $K_r$ dependence of $ZT$ in the three cases, as shown in Fig.~\ref{fig3} (e,f,g). Although, peak values of $ZT$ decrease with increasing $K_r$ in all three cases, the $se$ case has the lowest decreasing rate and still retain $ZT$ as large as 1.0 as $K_r$ increases up to 1.0 $K_0$. Meanwhile, compared to the blue shift of the $ZT$ peak position in both $ne$ and $de$ cases, an opposite red shift is found in the $se$ case, which once again renders the $se$ case a more preferable TE candidate. $K_r$ = 0.1$K_0$ is used from now on throughout this manuscript.

\begin{figure*}
\begin{center}\scalebox{0.42}{\includegraphics{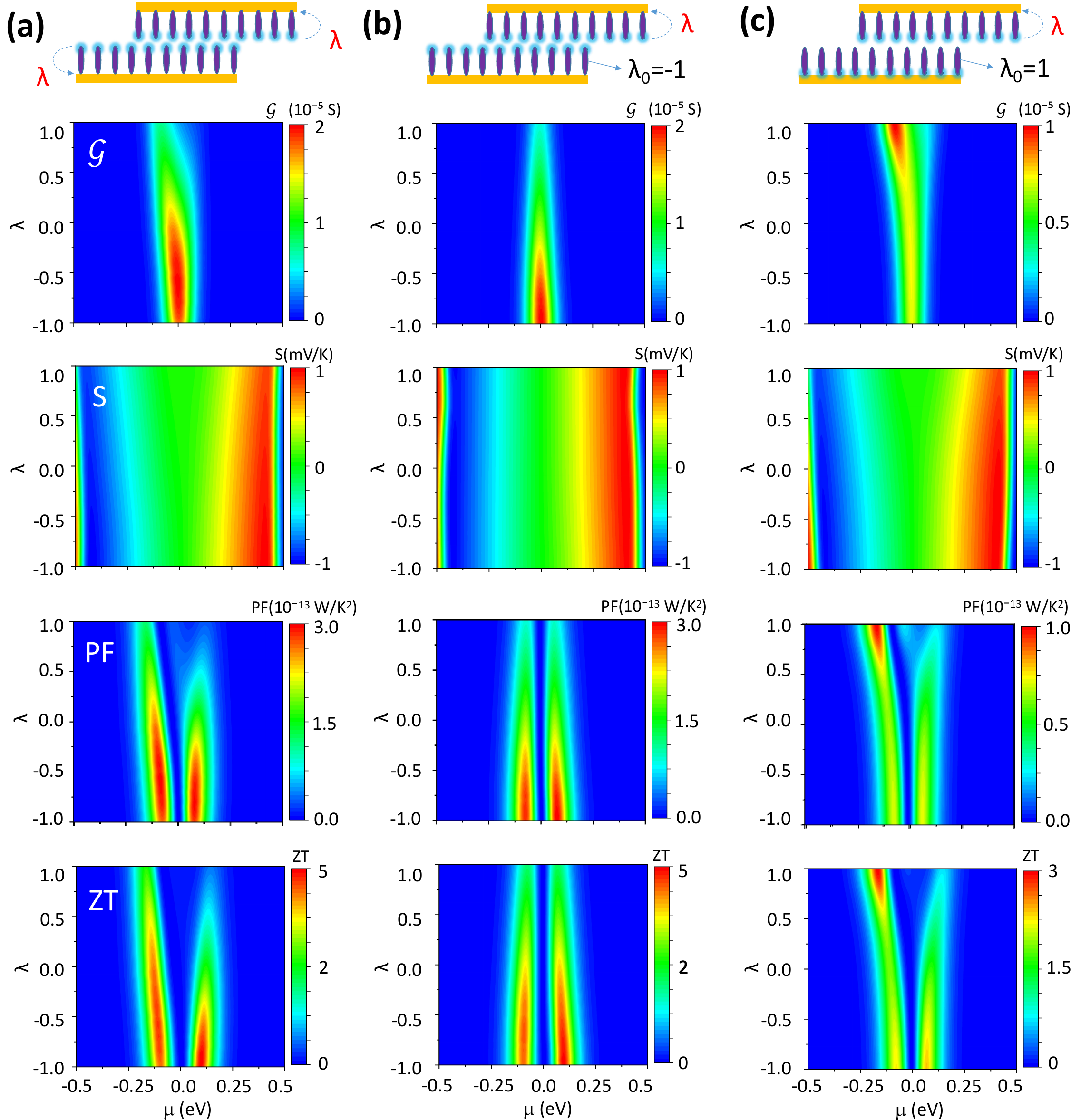}}
\caption{The impact of end states shift on thermoelectric properties of SSH SMD in the $se$ case. Shift of the end states from head to tail is realized via the topological phase transition of SSH chains, \textit{i.e.} by tuning the $\lambda$ of the electrode as indicated by the blue dashed arrow. From top to bottom are the $\lambda$ and $\mu$ dependent electrical conductance $\mathcal{G}$, Seebeck coefficient $S$, power factor $PF$, and $ZT$ value. From left to right, (a) represent the synchronous shift of the end states of both top and bottom electrodes, (b) the shift of only one electrode, (c) the shift of only one electrode, but the end states in another electrode anchored on the tail. Temperature at 300 K. $K_r$ = $0.1K_0$. The power factor is calculated via $PF=\mathcal{G} S^2$. 
\label{fig4}}
\end{center}
\end{figure*}

Now we study the optimization conditions by tuning the inter SSH chain hopping $t_h$ and $t_t$. Figure~\ref{fig3} (h-j) show maximum $ZT_{max}$ as a function of $t_h$ and $t_t$ for the three cases. The red zone indicates that $ZT_{max}$ can reach 3.0 in a wide range of $t_h$ and $t_t$ from 0 to 1.0$t_0$. In the red region, an approximate linear dependence of $t_h$ on $t_t$ is evident, and a ratio between $t_h$ and $t_t$, \textit{i.e.} $r_m = t_h/t_t$, will take some discrete numbers, emphasizing the importance of simultaneous tuning of the two hopping parameters. However, compared to the $ne$ (Fig.~\ref{fig3}(h)) and $de$ (Fig.~\ref{fig3}(j)) cases, which have a relatively wider red region, the $se$ case (Fig.~\ref{fig3}(j)) has $ZT_{max}$ confined in the regime of $r_m \leq 1.0$ or $t_h \geq 0.3 t_0$. This suggests an insensitivity of TE to $t_h$ in the $se$ case. In particular, when looking into the white dashed box area of Fig.~\ref{fig3}(j) and zooming in, one can see that the largest $ZT_{max}$ lies in a region centered by $t_h = 0.01t_0$ and $t_t = 0.12t_0$. This makes sense, since $t_h$ denotes hopping in the non-covalent connection and should be suppressed to obstruct phonon transport. Overall, good TE performance of the SSH SMD is expected at $t_h = 0.01t_0$, $t_t = 0.12t_0$ in the $se$ (single end state) case.


\begin{figure*}
\begin{center}\scalebox{0.42}{\includegraphics{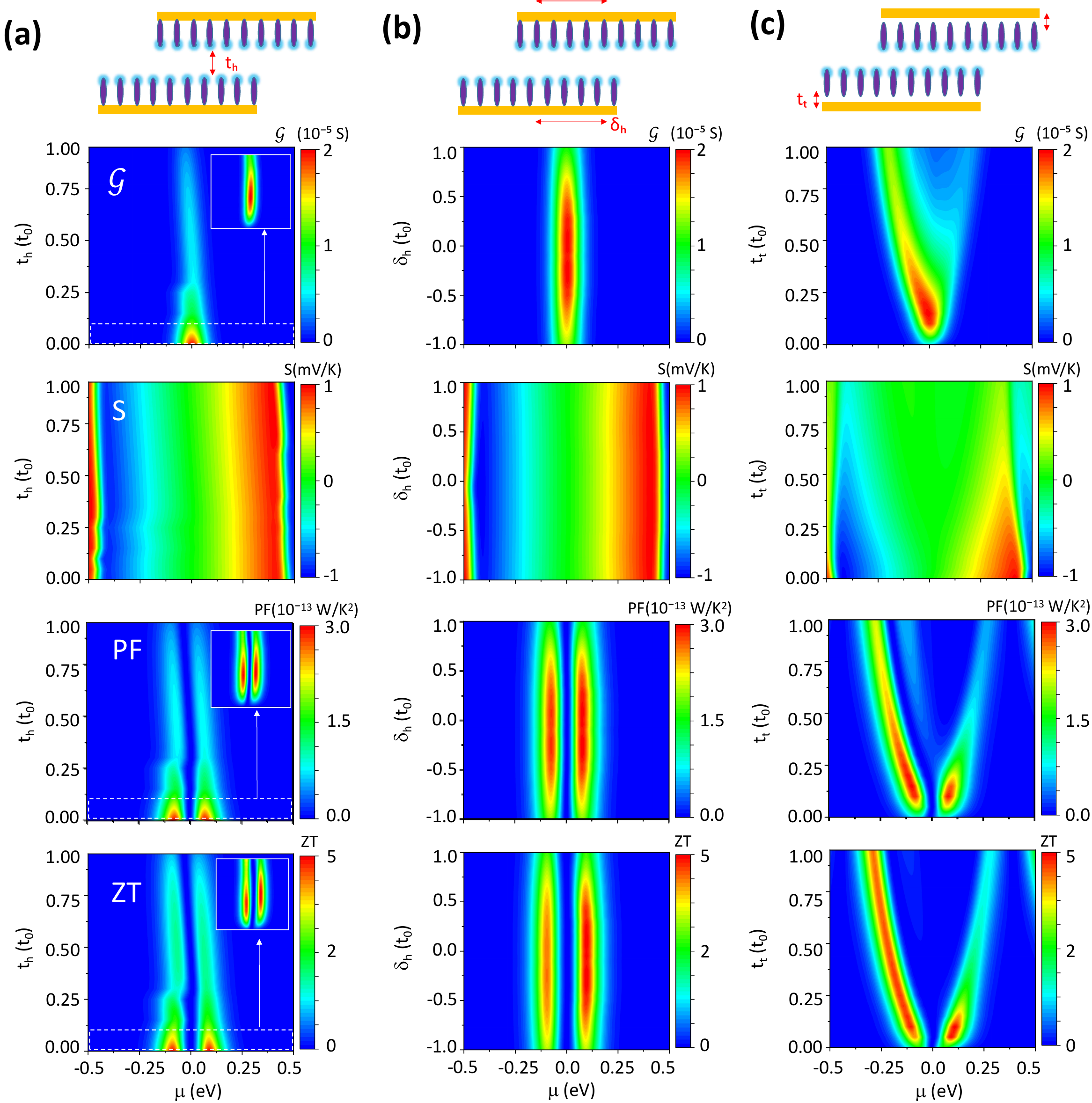}}
\caption{The impact of end states shift on thermoelectric properties of SSH SMD in the $se$ case. The shift is realized by three means: (a) vertical displacement between top and bottom leads to tune $t_h$, (b) horizontal displacement to tune $\delta_h$, and (c) vertical displacement between tail and lead to tune $t_t$. From top to bottom are tuning parameter ($t_h$, $\delta_h$, $t_t$) and $\mu$ dependent electrical conductance $\mathcal{G}$, Seebeck coefficient $S$, power factor $PF$, and $ZT$ value. Temperature is set at 300 K. The power factor is calculated via $PF=\mathcal{G} S^2$. $K_r$ = $0.1K_0$. The insets are zoomed in from the dashed square area with $t_h/t_0$ = [0, 0.02].
\label{fig5}}
\end{center}
\end{figure*}

\begin{figure*}
\begin{center}\scalebox{0.43}{\includegraphics{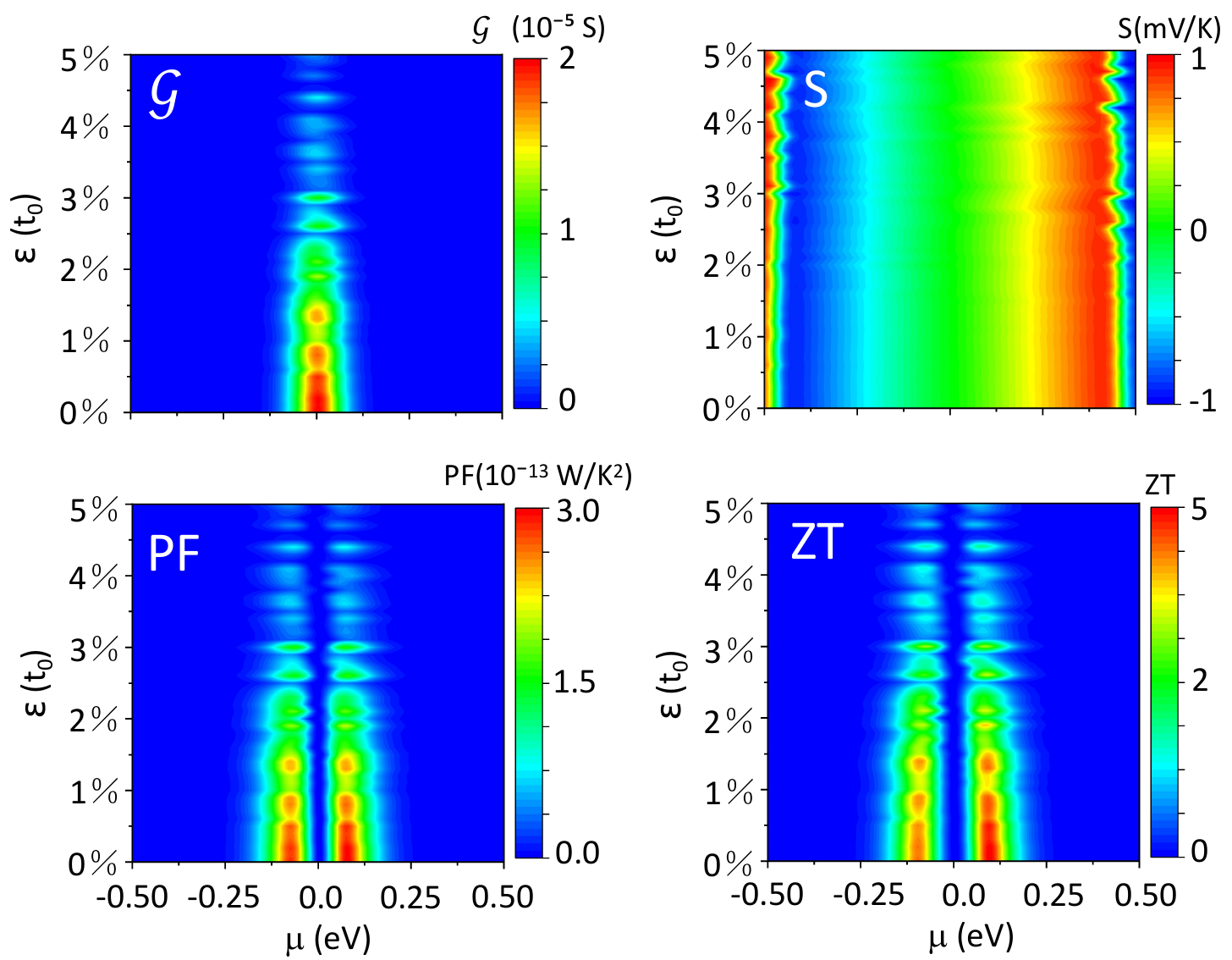}}
\caption{The disorder effect on the thermoelectric properties of the system in the $se$ case. The disorder perturbation is introduced by disordered on-site energy of $\varepsilon$ in the unit of $|t_0|$ (=3.002 eV). $K_r$ = 0.1 $K_0$. T = 300 K.
\label{sta}}
\end{center}
\end{figure*}


\subsection{Thermoelectric stability against perturbations}

From now on, we focus on the $se$ case and proceed to explore thermoelectric stability of the SSH SMD system against external perturbations of various kinds (i.e., end state shift, structural change, disorder, etc.), which are too ubiquitous in real devices to neglect. 


First, we investigate the impact of the end states shift. The shift of end states from one end to the other is simulated by continuously tuning the $\lambda$ from -1 to +1 in each SSH chain. As defined in the method section, the negative and positive $\lambda$ denote topological non-trivial and trivial states, respectively. In Fig.~\ref{fig4}, we present the $\lambda$-dependent TE properties ($\mathcal{G}$, $S$, $PF$ and $ZT$ from top to bottom) of SSH SMD. Fig.~\ref{fig4} (a) represents both end states shifting from head-to-head to tail-to-tail synchronously. Fig.~\ref{fig4} (b,c) represents one end state shifting from head to tail and the other anchored at head ($\lambda_0$ = -1) or tail ($\lambda_0$ = +1). Basically, the final configuration of (a) is also the final configuration of (c), the final configuration of (b) is the beginning of (c).   

From the TE properties in Fig.~\ref{fig4}, the best performance occurs in the head-to-head scenario with $\lambda \in [-1.0, 0.0)$ in (a) and (b), and the worst in the head-to-tail scenario with $\lambda_0$ = -1 and $\lambda \in [-1.0, 0.0)$ in (c). Fig.~\ref{fig4} (c) also shows that the worst scenario can be improved even by quenching all the topological end states (i.e., both end states at tail). Among the physical parameters concerned here, $S$ is much less sensitive to $\lambda$ in all the scenarios of (a-c). The $\lambda$ dependency of $PF$ and $ZT$ is mainly due to the switch effect of $\mathcal{G}$. The ratio of maximal $\mathcal{G}_{max}$ to minimal $\mathcal{G}_{min}$, which represents the TE degradation, in the vicinity of $\mu$= 0 is around 2.0, 4.0, and 2.0 in Fig.~\ref{fig4} (a), (b) and (c), respectively. To minimize TE degradation, it is, therefore, sensible to tune the end states symmetrically, such as in (a), not in (b).

The location of the end states is related to the chemical environment for the ends of the SSH chains, for instance, that of polyacetylene is related to the index of hydrogen deficiency of its terminal carbon atom, which can be tuned experimentally by end-capping engineering~\cite{ece}.

Second, we introduce other perturbations due to structural changes. Such changes can be vertical relative displacement, horizontal sliding and vertical displacement between tail and lead, as schematized on top of  Fig.~\ref{fig5} (a), (b) and (c), respectively. These changes can be achieved by varying the inter-SSH-chain hopping $t_h$, inducing inhomogeneity $\delta_h$ in head-to-head hopping, and tuning $t_t$. In more detail, $\delta_h=t_{h2}-t_{h1}$, representing the difference between two neighboring $t_h$.

As can be observed in Figs.~\ref{fig5} (a)-(c), transport property is more sensitive to $t_h$ than to both $t_t$ and $\delta_h$. Electrical conductance $\mathcal{G}$ changes more substantially than Seebeck coefficient, as $t_h$ and $t_t$ decrease from 1.0$t_0$, and thus, $PF$ and $ZT$ follows the behavior of $\mathcal{G}$. From Fig.~\ref{fig5} (a), it seems detrimental to have $t_h$ increased and deviated much from the optimal 0.01$t_0$, for example, by pushing two leads much closer to each other. For comparison, horizontal sliding in Fig.~\ref{fig5} (b) does not cause much harm to the device TE properties, and peak performance can be maintained in a wide range of $\delta_h$. Change of vertical attachment between tail and lead in Fig.~\ref{fig5} (c) is more complicated. Even though it can retain a good TE performance in a wide range of $t_t$, but with some constraints, for instance, electron doping is more preferable than hole doping and doping density needs to increase with increasing $t_t$, which obviously increases the complexity in control. So to maintain peak TE performance, vertical displacement in the SSH SMD, such as pushing and squeezing, should be restrained, once the optimal conditions are located. But not much constraints should be imposed on sliding.

Third, disorder as an intrinsic perturbation to the system is introduced and studied. To simulate the disorder effect, on-site energy of $\varepsilon = 0 \sim 5\%|t_0|$ is adopted, as shown in Fig.~\ref{sta}. Once again, $S$ shows a much better tolerance to disorder than $\mathcal{G}$. As $\varepsilon$ goes above 2$\%|t_0|$, areas of of $\mathcal{G}$, $PF$ and $ZT$ peak value become fuzzy and blurry. Disorder has to be less than this threshold value. 

\section{Summary}
In this work, the TE properties of an SMD made up of SSH chains have been theoretically investigated. 
The system can be categorized into three cases with different numbers of topological end states of unit SSH chain, \textit{i.e.} the $ne$, $de$ and $se$ cases. The $ne$ case possesses a giant gap, while the $de$ case has a narrow gap, and importantly, the $se$ case emerges like a single quantum dot. 
Decent TE performance of the system with maximum $ZT$ value up to 5.0 can be achieved in the $se$ case.  
By investigating the dependence on the inter SSH chain hopping energy, we found that TE performance can be optimized when the non-covalent hopping is weak to attenuate phonon transport. TE stability is also explored in terms of end state shift, structural change and disorders. 
Vertical displacement should be hindered but horizontal sliding is more tolerable. Disorder of on-site perturbation should be kept less than 2.5 $\%|t_0|$. In a word, the presence of topological end states makes the SSH SMD a good TE device for potential applications of energy conversion. 

\section*{Acknowledgments}
This project is supported by the National Natural Science Foundation of China 52031014, the Strategic Priority Research Program of the Chinese Academy of Sciences (Grant No. XDB0460000) and the National Key R\&D Program of China (2022YFA1203900). The simulation work was carried out at National Supercomputer Center in Tianjin, China, and the calculations were performed on TianHe-1(A). 

\bigskip


\end{document}